# Novel Tungsten Triboride

Yongcheng Liang[1], Xun Yuan[2], et al.

[1]*College of Engineering Science and Technology, Shanghai Ocean University, Shanghai 201306, China*

[2]*State Key Laboratory of High Performance Ceramics and Superfine Microstructures, Shanghai Institute of Ceramics, Chinese Academy of Sciences, Shanghai 200050, China*

A highly stable tungsten triboride with the Pearson symbol $hP24$ ($hP24$-$WB_3$) is identified by using density functional theory calculations. This new structure can be derived from the well-known $hP3$ configuration by removing one third of tungsten atoms systematically so that the remaining tungsten atoms form three cycled layers of open hexagons with each a layer displaced by one atom. Such a porous and metallic system has an unexpectedly high Vickers hardness of 38.3 GPa. It is revealed that a three-dimensional covalent framework composed of hexagonal boron planes interconnected with strong zigzag W-B bonds is responsible for its unusual high hardness.

The synthesis and prediction of new transition-metal borides (TMBs) possessing exceptional properties are the subject of an intense research activity since their unique characteristics confer a considerable technological interest to them. For example, following the discovery of a famous superconductor $MgB_2$ [1] that illustrates metallicity and covalency to be key ingredients for phonon-mediated superconductivity [2], a few TMBs such as $FeB_4$ [3], $YB_4$ [4] and $MoB_4$ [5] were found to be display appealing superconductivity. Strikingly, a large $TMB_6$ family attracts much attentions for they exhibit valence fluctuations ($SmB_6$) and Kondo ($CeB_6$) effects or have low work functions and low volatility at high temperature ($LaB_6$, $CeB_6$) allowing their use as high performance thermionic emitters [6]. Moreover, $FeB_2$ was predicted to be the first metal diboride semiconductor [3]. Recently, the boron incorporation into TMs forming dense TMBs ensures high valence electron density and strong covalent bonding resulting in high hardness [7]. Applying this principle, a series of promising superhard TMBs were experimentally synthesized or theoretically predicted [8]. In the following, we demonstrate that a novel TMB with unusual behaviors can be found in such a well-known and accessible binary system as W-B.

A great interest is currently focused on tungsten borides, not only because they can form rich phases, but also because of their high potentials to serve as relatively inexpensive superhard materials. The early established $hP14$-$W_2B_5$ [9] was recently identified as $hP12$-$WB_2$ [10], which has a maximum measured hardness of 42.2 GPa [11]. First-principles calculations suggested that $hP6$-$WB_2$ was energetically more stable than the experimentally reported $hP3$-$WB_2$ [12] and could have the high hardness with a value of 43.8-47.4 GPa [13, 14]. It was once accepted that the claimed $hP20$-$WB_4$ [15] was an inexpensive superhard material (Vickers hardness 43.3-46.2 GPa under the load of 0.49 N) [16-19]. However, this long-assumed $hP20$-$WB_4$ has recently identified as $hP16$-$WB_3$ from the thermodynamic stability [20], mechanical properties [21] and XRD spectra comparisons [22, 23]. These works have solved much controversy on crystal structures and mechanical properties of the W-B system but have not systematically explored new stable tungsten borides.

In this Letter, we present a comprehensive study of the phase stability, crystal structure and electronic properties for the W-B system. We find a brand new $hP24$-$WB_3$ that can be obtained from the well-known $hP3$-$WB_2$ by removing one third of tungsten atoms systematically. This noncompact and conductive $hP24$-$WB_3$ not only is the highest boride of tungsten with the thermodynamic stability but also has an unexpectedly high hardness of 38.3 GPa. Furthermore, we reveal that such an anomalous hardness has its electronic origin. This newly identified compound strongly challenges the common strategy only pursuing dense TMBs for superhard materials.

The calculations were carried out using the density functional theory within the generalized gradient approximation (GGA) [24] as implemented in the Vienna *ab initio* simulation package (VASP) [25]. The chosen cutoff energy of 500 eV and dense *k*-point meshes ensure numerical convergence of energy differences to typically ~1 meV/atom. All considered structures were optimized with respect to both lattice parameters and atomic positions. Their formation energies $\Delta E$ were calculated according to $\Delta E = E(W_{1-x}B_x) - (1-x)E(W) - xE(B)$ at the temperature $T$=0 K, and bcc-W and α-B were adopted for the respective ground state. For phases with minute energy differences, the temperature effect was also included through phonon corrections to Gibbs free energy $G(T)$ using a finite displacement method.

We first characterize the structural, dynamical, electronic, and mechanical properties of the new $hP24$-$WB_3$ phase. Its crystal structure with the space group $R$-$3m$ (No.166) is plotted in Fig. 1(a). At the equilibrium condition, the hexagonal lattice parameters are $a$=5.213 Å and $c$=9.409 Å with the tungsten and boron atoms occupying the Wyckoff positions *6c* (0, 0, 0.166) and *18f*

(0.665, 0, 0), respectively. This structure is stacked alternately in terms of planar boron and tungsten layers. Specially, there exist the holes, which are large enough to accommodate at least two boron atoms, in this structure centered at the sites $3b$ (0, 0, 0.5). It is known that phonons are a strict measure for structural stability and thus we calculate its phonon dispersion curves. In Fig. 1(b), no negative frequencies are observed throughout the whole Brillioun zone, confirming dynamical stability of $hP24$-$WB_3$. From the electronic band structure in Fig. 1(c), we note that there are several bands crossing the Fermi level, indicating well metallic feature in $hP24$-$WB_3$.

To evaluate its mechanical stability, we calculate elastic constants of $hP24$-$WB_3$. The obtained elastic constants ($C_{11}$=659 GPa, $C_{12}$=90 GPa, $C_{13}$=182 GPa, $C_{33}$=472 GPa, $C_{44}$=278 GPa) satisfy the Born-Huang criterion [26] for a hexagonal crystal, implying that it is mechanically stable at ambient conditions. Interestingly, we predict that $hP24$-$WB_3$ has the relatively high bulk modulus $B$ (299 GPa) and Young's modulus $E$ (587 GPa). Moreover, it also exhibits exceptionally high shear modulus $G$ (251 GPa) and very low Poisson's ratio $v$ (0.172). According to the recently proposed hardness model [27], we predict its Vickers hardness $H$ to be an unexpectedly high value of 38.3 GPa. These values are very close to that of the recently identified $hP16$-$WB_3$ ($B$=295 GPa, $E$=588 GPa, $G$=252 GPa, $v$=0.168, $H$=39.2 GPa) [21] and even match that of $ReB_2$ ($B$=356 GPa, $E$=691 GPa, $G$=293 GPa, $v$=0.181, $H$=41.2 GPa) [28]. Hence, our results show that $hP24$-$WB_3$ should be an incompressible and highly hard material.

As we know, most of the hard or superhard materials such as diamond and cBN are insulators or semiconductors with dense structures. It therefore is a bit surprising that a porous and metallic system is so hard. To elucidate the electronic origin responsible for such an anomalous hardness, we perform the calculations of difference charge density for $hP24$-$WB_3$. From Fig. 1(d), we can clearly see that the extremely strong B-B covalent bonds become evident and form hexagonal graphite-like meshes on the (001) plane. On the (110) plane, a large amount of charges piles up between the tungsten and boron atoms. This indicates that there is powerful bonding of W-B, creating well-defined zigzag directional chains along the $z$-direction. Therefore, the hexagonal covalent boron meshes interconnected with the zigzag directional chains compose a three-dimensional rigid framework. It is this three-dimensional rigid network that plays a key role in the unusual high hardness of $hP24$-$WB_3$.

We now study the thermodynamic stability for the W-B system in order to confirm the viability of $hP24$-$WB_3$. The calculated $T$=0 K formation energies for fourteen candidate structure types are presented in Fig. 2(a). The $tI$12, $tI$16, $oC$8, $oP$8, $hP$4, $hP$3, $hR$18, $hP$12, $hP$6, $oP$6, $hP$16, $hR$24, $hP$10 and $oI$10 structures correspond to the $W_2B$, $WB$, $CrB$, $FeB$, $NiAs$, $AlB_2$, $MoB_2$, $WB_2$, $ReB_2$, $OsB_2$, $WB_3$, $WB_3$ (proposed), $WB_4$ and $CrB_4$ prototypes, respectively. First of all, we can see that the convex hull plotted (in green) through the $tI$12-$W_2B$, $tI$16-$WB$ and $hP$6-$WB_2$ ground states reproduces that of the recent theoretical work [14, 20], substantiating the reliability of the present calculations. At 1:1 composition, we can see that the low-temperature phase $tI$16-$WB$ is thermodynamically stable whereas the high-temperature phase $oC$8-$WB$ is metastable. However, the latter becomes more stable than the former at $T$=2113 K [see Fig. 2(b)], in good agreement with the experimental observation [29]. Although the hypothetical $oP$8-$WB$ is more stable than the two experimentally observed phases ($tI$16-$WB$ and $oC$8-$WB$) at $T$>2879 K, a phonon dispersion calculation shows dynamical instability of $oP$8-$WB$ with a small imaginary frequency for a $\Gamma$-point phonon. In addition, the assumed $hP$4-$WB$ has the extremely high relative formation energy and should be thermodynamically unstable.

At 1:2 composition, the experimentally reported $hP$3-$WB_2$ [12] has by far the highest relative formation energy (-83 meV/atom) and thus it seems unlikely that it occurs at ambient condition. Unexpectedly, a phonon dispersion calculation shows that it is also dynamically unstable with imaginary frequencies. Interestingly, $hP$6-$WB_2$ has the lowest formation energy (-366 meV/atom) and becomes a thermodynamic ground state, which accords with the recent first-principles calculations [13, 14, 20]. At the same time, we notice that $hP$12-$WB_2$, $hR$18-$WB_2$ and $oP$6-$WB_2$ are metastable within 27 meV/atom. Figure 2(c) reflects the difference in the vibrational properties of $hP$12, $hR$18, $oP$6 and $hP$6 and explains why the only $hP$12-$WB_2$ was observed during the high temperature synthesis [10].

At 1:3 composition, the recently identified $hP$16-$WB_3$ [20] breaks the $\alpha$-B and $hP$6-$WB_2$ tieline by 15 meV/atom and can be a viable phase under ambient condition. Surprisingly, we find that the proposed $hR$24-$WB_3$ (-305 meV/atom) has the lower formation energy than the metastable $hP$16-$WB_3$ (-289 meV/atom) and becomes a thermodynamic ground state. Moreover, our calculated Gibbs free energies including the temperature effect show that this new phase remains stable up to $T$=3000 K [see Fig. 2(d)]. At 1:4 composition, the recently predicted $hP$10-$WB_4$ [23] is found to be metastable with respect to $\alpha$-B and $hR$24-$WB_3$ whereas the hypothetical $oI$10-$WB_4$ may be unstable due to its relatively high formation energy. Therefore, our results suggest that the newly identified $hP24$-WB$_3$ is the most thermodynamically stable phase with ultimate boron content for W-B compounds.

However, it is a bit puzzling that the experimentally reported $hP3$-WB$_2$ exhibits the thermodynamic and dynamical instability whereas the theoretically predicted $hP6$-WB$_2$ and $hR24$-WB$_3$ (the experimentally synthesized $hP12$-WB$_2$ and $hP16$-WB$_3$) are thermodynamically stable (metastable). To clarify the underlying origin, we will systematically discuss their structural relations and electronic nature.

The well-known structure of $hP3$-WB$_2$ structure can be described with the stacking sequence AHAH in the $c$ direction. The H layer is a planar, graphite-like mesh of boron atoms, while the tungsten atoms in the A layers occupy positions directly above and below the open centers of the boron hexagons, thus forming a close-packed metal layer (see A and H in Fig. 3). In this configuration, the tungsten atoms coordinate twelve boron atoms arranged in the two regular hexagons and have six tungsten neighbors in the same layer and two tungsten neighbors in the $c$ direction [see (I) in Fig. 3]. With increasing filling of the TM $d$ shell, this structure becomes increasingly less stable [30]. We clearly see from Fig. 4(a) and (I) that the energy range from -2.5 eV to 1.5 eV is dominated by the W-5$d_z^2$ states and the $c$-axis W-W antibonding interactions are formed. The thermodynamic and dynamical instability of $hP3$-WB$_2$ is mainly attributed to its high density of states (DOS) at the Fermi level and a substantial filling of the strong antibonding states (-2.5–0 eV). In order to enhance the structural stability, this fierce antibonding interaction should be relieved by the formation of complex, but closely related structures. It is found that two favorable pathways (staggering tungsten atoms and removing tungsten atoms) can reduce the antibonding interaction.

The first type of structural modifications ($hP12$-WB$_2$ and $hP6$-WB$_2$) is the result of a translation of tungsten atoms in $hP3$-WB$_2$. We notice that $hP12$-WB$_2$ can be obtained from the $hP3$-WB$_2$ by translating double tungsten layers with respect to each other so as to stagger the tungsten atoms along the $c$ direction, while puckering boron layers between the staggered tungsten layers so that the close W-B distances remain. Consequently, $hP12$-WB$_2$ is stacked by the sequence AHAK'BHBK', where the B layer is the tungsten layer shifted by ($a$/3, 2$a$/3) with reference to A and K' signifies the puckered boron layer (see A, B, H and K' in Fig. 3). Compared with $hP3$-$WB_2$, the tungsten environment in $hP12$-$WB_2$ is evidently different. Half of boron atoms remain a regular hexagon while the other half of boron atoms are distorted into a trigonal pyramid in place of a tungsten atom in the $c$ direction [see (II) in Fig. 3]. The calculated DOS in $hP12$-$WB_2$ shows a higher hybridization of the B-2$p$ and W-5$d$ states with the antibonding interaction of the W $5d_z^2$ orbitals now lying nearly above the Fermi level. Moreover, this distortion results in a dramatic reduction of the high DOS at the Fermi level [see (b) and (II) in Fig. 4]. Hence, the formation energy drastically decreases from $hP3$-$WB_2$ to $hP12$-$WB_2$. Similarly, the hypothetical $hR18$-$WB_2$ (AHAK'BHBK'CHCK') and $oP6$-$WB_2$ (distorted AK'AK') have related structures, and thus they exhibit the low formation energy. However, these distortions are incomplete. When all tungsten layers are translated with respect to one another and all boron layers are puckered, this complete distortion gives $hP6$-$WB_2$ that has the stacking sequence AK'BK' (see A, B and K' in Fig. 3). Around each tungsten atom, two regular hexagons of boron atoms in $hP3$-$WB_2$ are distorted into two trigonal pyramids in $hP6$-$WB_2$ and no direct W-W contacts occur along the $c$ direction [see (III) in Fig. 3]. As shown in Fig. 4(c) and (III), the Fermi level of $hP6$-$WB_2$ just locates at a distinct minimum in its DOS curve and there is a strong hybridization of the B-2$p$ and W-5$d$ states, especially for the W $5d_z^2$ states. It is therefore natural that $hP6$-$WB_2$ has the lowest formation energy among tungsten diborides and becomes a thermodynamic ground state.

The second type of structural modifications ($hP16$-$WB_3$ and $hR24$-$WB_3$) is the result of a removal of tungsten atoms in $hP3$-$WB_2$. We surprisingly find that $hP16$-$WB_3$ ($hR24$-$WB_3$) can be derived from $hP3$-$WB_2$ by removing one third of the tungsten atoms systematically so that the remaining tungsten atoms form two (three) cycled layers of open hexagons with each a layer displaced by one atom. If A' denote the close-packed tungsten layers A with one third of the tungsten atoms removed, B' and C' are the tungsten layer shifted by one atom and two atoms with reference to A', respectively (see A', B', C' in Fig. 3). Accordingly, $hP16$-$WB_3$ and $hR24$-$WB_3$ can be regarded as the stacking sequence A'HB'H and A'HB'HC'H, respectively. This "defect" of the tungsten atoms in $hP16$-$WB_3$ and $hR24$-$WB_3$ proves to be a more favorable way of reducing the high DOS at the Fermi level. From Fig. 4(d) and (e), we can see that both Fermi levels lie in the pronounced distinct minimum in their DOS curves, respectively. At the same time, their electronic structures are governed by the strong hybridization between the W-5$d$ and B-2$p$ states and the powerful covalent bonds of W-B are formed. Thence, $hP16$-$WB_3$ and $hR24$-$WB_3$ achieve the very low energy of formation. However, the local environments of tungsten atoms are different for $hP$16-WB$_3$ and $hR$24-WB$_3$. For $hP$16-WB$_3$, half of tungsten atoms have three tungsten neighbors belonging to the same metal layer whereas the other half have five tungsten neighbors in a bipyramidal arrangement [see (IV) and (V) in Fig. 3]. Although the former greatly relieve the W-W antibonding interaction along the $c$ direction, the latter still have this interaction. Our calculated $5d$ projected DOS of two inequivalent tungsten atoms confirm this point [see see (IV) and (V) in Fig. 4]. For $hR$24-WB$_3$, each tungsten atom has four tungsten neighbors in a pyramidal arrangement [see (VI) in Fig. 3]. This local coordination further relieves the $c$-axis W-W interaction [see (VI) in Fig.4], and thus $hR$24-WB$_3$ becomes a thermodynamic ground state.

In summary, we have systematically studied the phase stability, crystal structure and electronic properties for the W-B system by using the first-principles calculations. It is identified that the novel $hR$24-WB$_3$ can be derived from the known $hP$3-WB$_2$ by removing one third of tungsten atoms systematically so that the remaining tungsten atoms form three cycled layers of open hexagons with each a layer displaced by one atom. To the best of our knowledge, this never observed $hR$24-WB$_3$ phase is the most thermodynamically stable phase with ultimate boron content for W-B compounds so far discovered. Our results show this porous and metallic system has an unexpectedly high hardness of 38.3 GPa. Moreover, we reveal that a three-dimensional covalent framework composed of hexagonal boron planes interconnected with strong zigzag W-B bonds is responsible for its unusual high hardness. The present work challenges the common opinion only searching for dense TMBs for superhard materials and highlights the importance of thermodynamic stability and three-dimensional network in design of intrinsically hard materials.

This work is financially supported by the NSFC (No. 51072213), State Oceanic Administration (No.SHME2011GD01), the Local Colleges Faculty Construction of Shanghai MSTC (No. 11160501000), and the Innovation Program of Shanghai MEC (No.11ZZ147).

[1] J. Nagamatsu, N. Nakagawa, T. Muranaka, Y. Zenitani, J. Akimitsu, Nature **410**, 63 (2001).

[2] H. J. Choi, D. Roundy, H. Sun, M. L. Cohen, and S. G. Louie, Nature **418**, 758 (2002).

[3] A. N. Kolmogorov, S. Shah, E. R. Margine, A. F. Bialon, T. Hammerschmidt, and R.Drautz, Phys. Rev. Lett. **105**, 217003 (2010).

[4] Y. Xu, L. Zhang, T. Cui, Y. Li, Y. Xie, W. Yu, Y. Ma, and G. Zou, Phys. Rev. B **76**, 214103 (2007).

[5] J. W. Simonson, D. Wu, S. J. Poon, and S. A. Wolf, J. Supercond. Nov. Magn. **23**, 417 (2010).

[6] A. N. Kolmogorov *et al*., Phys. Rev. Lett. 109, 075501 (2012); H. Zhang, *et al*., J. Am. Chem. Soc. **127**, 2862 (2005); X. H. Ji *et al*., Prog. Solid State Chem. **39**, 51 (2011).

[7] R. B. Kaner *et al*., Science **308**, 1268 (2005); H. -Y. Chung *et al*., Science **316**, 436 (2007); R.W. Cumberland *et al*., J. Am. Chem. Soc. **127**, 7264 (2005).

[8] H. Niu *et al*., Phys. Rev. B **85**, 144116 (2012); J. V. Rau and A. Latini, Chem. Mater. **21**, 1407 (2009); H. Y. Gou *et al*., Appl. Phys. Lett. **100**, 111907 (2012); M. Zhang *et al*., J. Phys. Chem. C **114**, 6722 (2010); Q. Wang *et al*., J. Phys. Chem. C **115**, 19910 (2011); X. Meng *et al*., J. Appl. Phys. **111**, 112616 (2012).

[9] R. Kiessing, Acta Chem. Scand. **1**, 893 (1947).

[10] M. Frotscher, W. Klein, J. Bauer, C. M. Fang, J. F. Halet, A. Senyshyn, C. Baehtz, and B. Albert, Z. Anorg. Allg. Chem. **633**, 2626 (2007).

[11] J. V. Rau, A. Latini, R. Teghil, A. D. Bonis, M. Fosca, R. Caminiti, and V. R. Albertini, ACS Appl. Mater. Interfaces **3**, 3738 (2011).

[12] H. P. Woods, F. E. Wawner, and B. G. Fox, Science **151**, 75 (1966).

[13] X. Q. Chen, C. L. Fu, M. Krcmar, and G. S. Painter, Phys. Rev. Lett. **100**, 196403 (2008).

[14] E. Zhao, J. Meng, Y. Ma, and Z. Wu, Phys. Chem. Chem. Phys. **12**, 13158 (2010).

[15] P. A. Romans and M. P. Krug, Acta Cryst. **20**, 313 (1966).

[16] Q. Gu, G. Krauss, and W. Steurer, Adv. Mater. **20**, 3620 (2008).

[17] R. Mohammadi, A. T. Lech, M. Xie, B. E. Weaver, M. T. Yeung, S. H. Tolbert, and R. B. Kaner, Proc. Natl. Acad. Sci. U.S.A **108**, 10958 (2011).

[18] M. Xie, R. Mohammadi, Z. Mao, M. Armenttrout, A. Kavner, K. B. Kaner, and S. H. Tolbert, Phys. Rev. B **85**, 064118 (2012).

[19] M. Wang, Y. Li, T. Cui, Y. Ma, and G. Zou, Appl. Phys. Lett. **93**, 101905 (2008).

[20] Y. Liang, X. Yuan, and W. Zhang, Phys. Rev. B **83**, 220102(R) (2011).

[21] Y. Liang, Z. Fu, X. Yuan, S. Wang, Z. Zhong, and W. Zhang, Europhys. Lett. **98**, 66004 (2012).

[22] R. F. Zhang, D. Legut, Z. J. Lin, Y. S. Zhao, H. K. Mao, and S. Veprek, Phys. Rev. Lett. **108**, 255502 (2012).

[23] H. Gou, Z. Li., L. Wang, J. Lian, and Y. Wang, AIP Adv. **2**, 012171 (2012).

[24] J. P. Perdew, K. Burke, and M. Ernzerhof, Phys. Rev. Lett. **77**, 3865 (1996).

[25] G. Kresse and J. Furthmuller, Phys. Rev. B **54**, 11169 (1996).

[26] M. Born and K. Huang, *Dynamic Theory of Crystal Lattice* (Oxford University Press, Oxford, 1954).

[27] H. Niu, J. Wang, X. Q. Chen, D. Li, Y. Li, P. Lazar, R. Podloucky, and A. N. Kolmogorov, Phys. Rev. B **85**, 144116 (2012).

[28] Y. Liang and B. Zhang, Phys. Rev. B **76**, 132101 (2007).

[29] H. Duschanek, and P. Rogl, J. Phase Equilibria **16**, 150 (1995).

[30] J. K. Burdett, E. Canadell, G. J. Miller, J. Am. Chem. Soc. **107**, 6561 (1986).

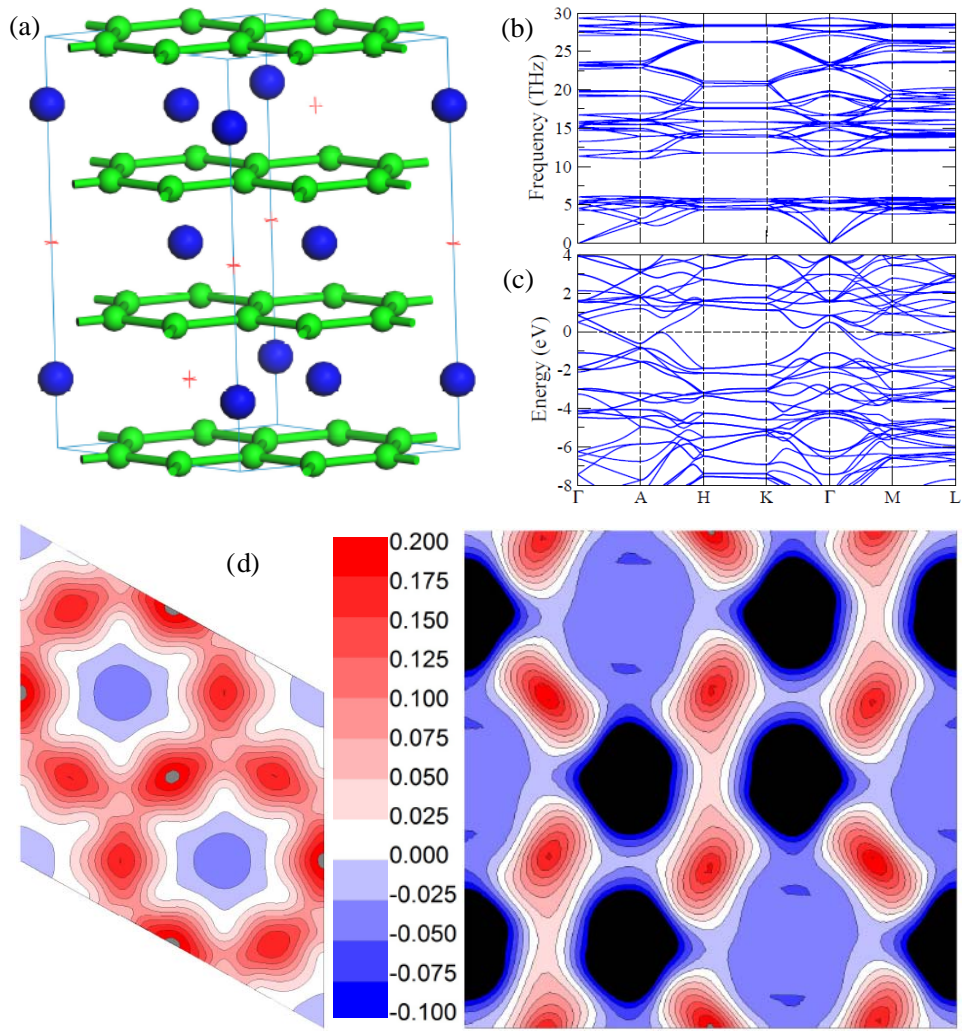

FIG. 1 (color online). Properties of $hP24$-$WB_3$ in $R$-$3m$ symmetry. (a) Crystal structure; (b) Calculated phonon dispersion curves; (c) Calculated electronic band structure; (d) Contours of the difference charge density on the (001) and (110) planes. In (a), the blue big and green small spheres represent the tungsten and boron atom, respectively. This $hP24$-$WB_3$ structure is porous and the center of each hole is shown by the red mark.

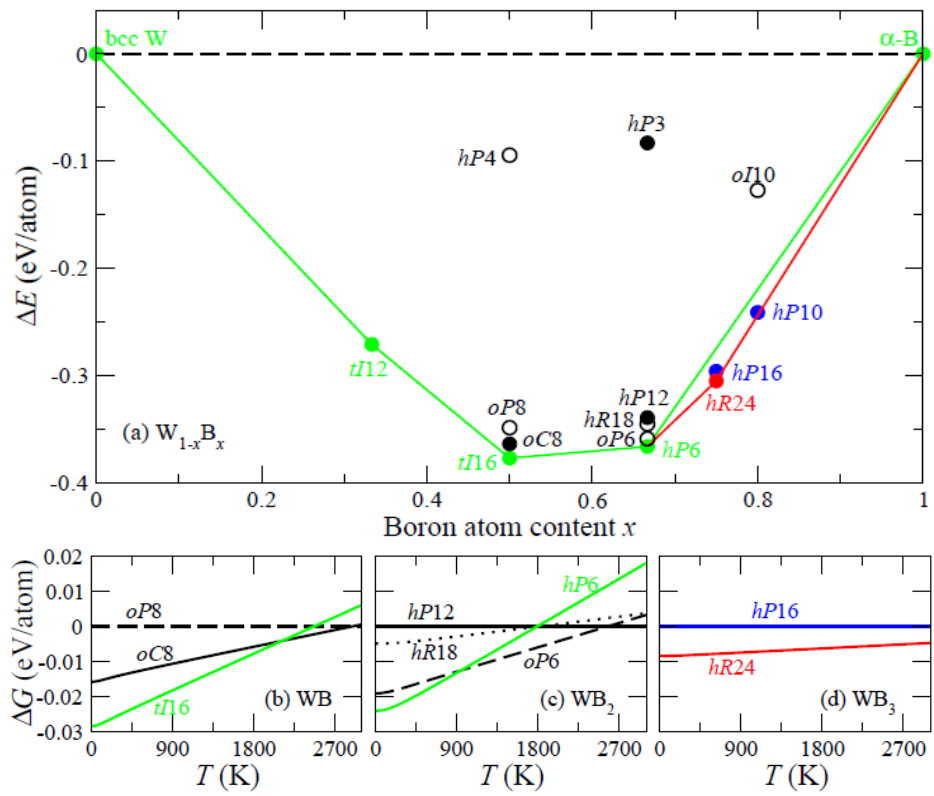

FIG. 2 (color online). Relative stability of the W-B system. (a) Formation energies $\Delta E$ of different composition phases at $T$=0 K; (b)-(d) Gibbs free energies $\Delta G$ of different structure phases with the same composition as a function of temperature $T$.

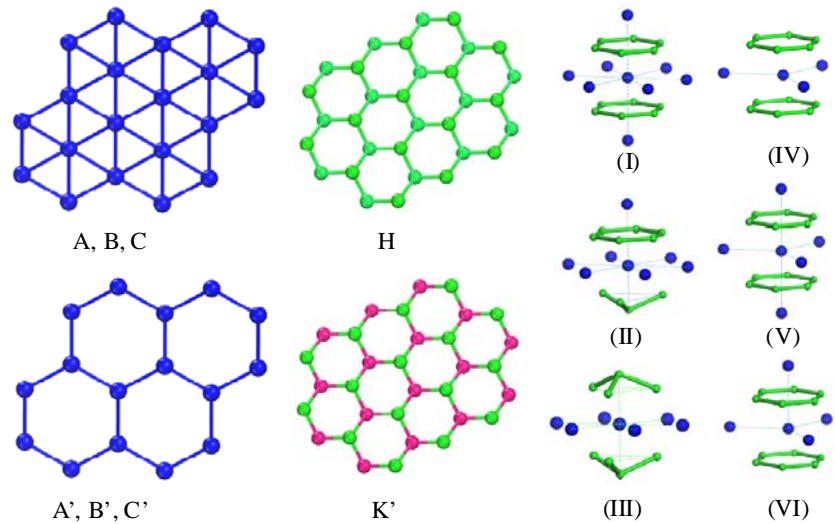

FIG. 3 (color online). Tungsten and boron layers and local coordination environments. Defect tungsten layers (A', B', C') can be derived from close-packed tungsten layers (A, B, C, respectively) by removing one third of tungsten atoms systematically. The boron layer H is planar while the boron layer K' is puckered. In K' layer, the red and green atoms are at different heights in the *c* direction. The six different local coordination environments (I)-(VI) of the tungsten atoms are shown. The blue and green (red) small spheres represent the tungsten and boron atom, respectively.

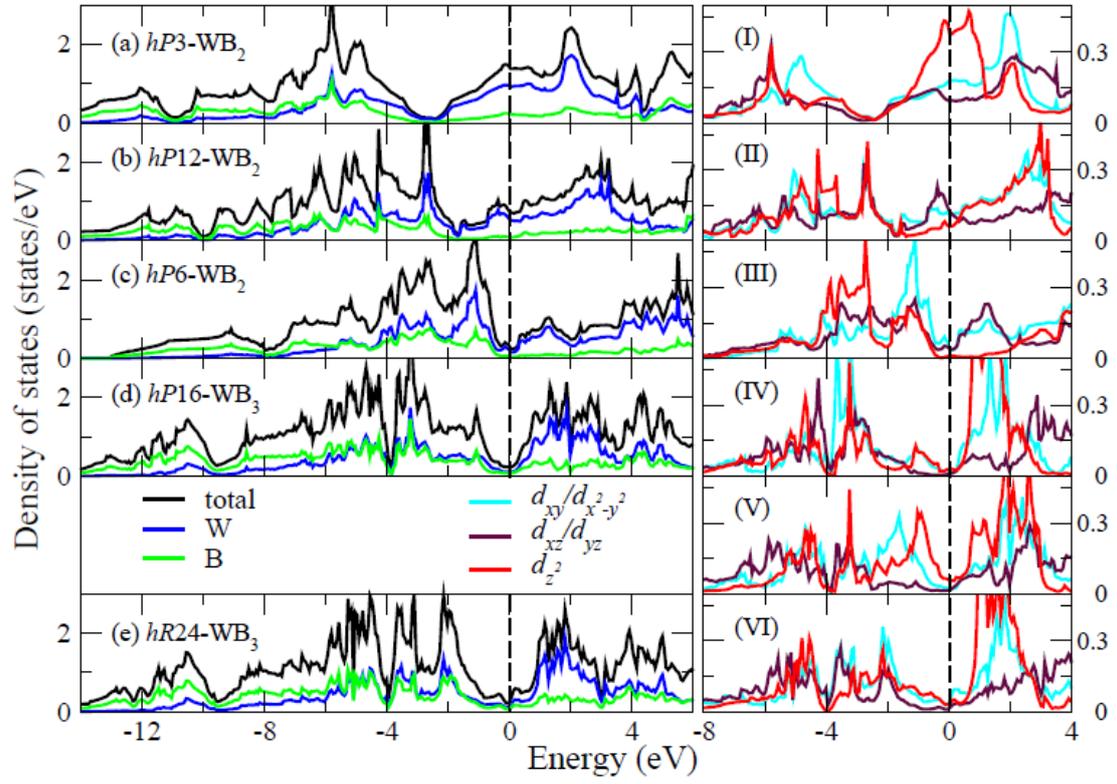

FIG. 4 (color online). Calculated density of states (DOS) for five tungsten borides and the W 5$d$-projected DOS for the corresponding six different coordination environments (I)-(VI) in Fig. 3.